\documentclass{elsart}

\usepackage{graphicx}
\usepackage{amssymb}

\begin{document}
\begin{frontmatter}
\title{Efficient Calculation of Energy Spectra Using Path Integrals\thanksref{MNTR}}
\thanks[MNTR]{Supported by the Ministry of Science and Environmental Protection of the Republic of Serbia through project No. 141035.}

\author{D. Stojiljkovi\' c\corauthref{daca}},
\ead{danica@phy.bg.ac.yu}
\corauth[daca]{Corresponding author.}
\author{A. Bogojevi\' c}, \and
\author{A. Bala\v z}

\address{Scientific Computing Laboratory, Institute of Physics\\
P. O. Box 57, 11001 Belgrade, Serbia}

\begin{abstract}
A newly developed method for systematically improving the convergence of path integrals for transition amplitudes, introduced in \textit{Phys. Rev. Lett.} 94 (2005) 180403, \textit{Phys. Rev.} \textbf{B} 72 (2005) 064302, \textit{Phys. Lett.} \textbf{A} 344 (2005) 84, and expectation values, introduced in \textit{Phys. Lett.} \textbf{A} 360 (2006) 217, is here applied to the efficient calculation of energy spectra. We show how the derived hierarchies of effective actions lead to substantial speedup of the standard path integral Monte Carlo evaluation of energy levels. The general results and the ensuing increase in efficiency of several orders of magnitude are shown using explicit Monte Carlo simulations of several distinct models.
\end{abstract}

\end{frontmatter}

\section{Introduction}
\label{sec:intro}

Feynman's path integrals \cite{feynmanhibbs,feynman} provide the general mathematical framework for dealing with quantum and statistical systems. The formalism has been successfully applied in generalizing the quantization procedure from the archetypical quantum mechanical problem of the dynamics of a single particle moving in one dimension, to more particles, more dimensions, as well as to more complicated objects such as fields, strings \cite{itzyksonzuber}, etc. Symmetries of physical systems can be more easily treated and applied in this formalism, since it gives a simple and natural setup for their use \cite{coleman}. Various approximation techniques are more easily derived within the framework of this formalism, and it has been successfully used for deriving non-perturbative results. The parallel application of this formalism in both high energy and condensed matter physics makes it an important general tool \cite{itzyksondrouffe,parisi}. The analytical and numerical approaches to path integrals have by now become central to the development of many other areas of physics, chemistry and materials science, as well as to the mathematics and finance \cite{barkerhenderson,pollockceperley,ceperley,kleinert}. In particular, general numerical approaches such as the path integral Monte Carlo method have made possible the treatment of a wealth of non-trivial and previously inaccessible models.

The key impediment to the development of the path integral formalism is a lack of complete understanding of the general mathematical properties of these objects. In numerical approaches limited analytical input generally translates into lower efficiency of employed algorithms. The best path generating algorithms, for example, are efficient precisely because they have built into them the kinematic consequences of the stochastic self-similarity of paths \cite{jaggedness}. A recent series of papers \cite{prl,prb,euler} has for this reason focused on the dynamical implications of stochastic self-similarity by studying the relation between path integral discretizations of different coarseness. This has resulted in a systematic analytical construction of a hierarchy of $N$-fold discretized effective actions $S^{(p)}_N$ labeled by a whole number $p$ and built up from the naively discretized action in the mid-point ordering prescription (corresponding to $p=1$). The level $p$ effective actions lead to discretized transition amplitudes and expectation values differing from the continuum limit by a term of order $1/N^p$.

In this paper we extend the applicability of the above method for improving the efficiency of path integral calculations to the evaluation of energy spectra. We show how the increased convergence of path integrals translates into the speedup in the numerical calculation of energy levels. Throughout the paper we present and comment on the Monte Carlo simulations conducted using the hierarchy of effective actions for the case of several different models including anharmonic oscillator, P\"oschl-Teller potential, and Morse potential. All the numerical simulations presented were done using Grid-adapted Monte Carlo code and were run on EGEE-II and SEE-GRID-2 infrastructure \cite{egee,seegrid}. The effective actions and the codes used can be found on our web site \cite{scl}.

\section{Partition Function and Energy Spectra}

The partition function is the central object in statistical mechanics. The path integral formalism gives us an elegant framework for calculating partition functions which can be used either for deriving analytical approximation techniques or for carrying out numerical evaluation. The starting point is the expression for the partition function in the coordinate basis,
\begin{eqnarray}
\label{partition}
Z(\beta)=\int_{-\infty}^{\infty}da\,A(a,a;\beta)\ ,
\end{eqnarray}
where $A(a,b;\beta)=\langle b|e^{-\beta\hat H}|a\rangle$ is the quantum mechanical transition amplitude for going from $a$ to $b$ in (Euclidean) time $\beta$. In the path integral formalism transition amplitudes are given as the $N\to\infty$ limit of the $(N-1)$-fold integral expression
\begin{equation}
\label{amplitudeN}
A_N(a,b;\beta)=\left(\frac{1}{2\pi\epsilon_N}\right)^{\frac{N}{2}}\int dq_1\cdots dq_{N-1}\,e^{-S_N}\ .
\end{equation}
$S_N$ is the naively discretized action of the theory, $\epsilon_N=\beta/N$ the discrete time step. For the physical models that we consider the action is of the form
\begin{equation}
\label{action}
S=\int_0^\beta dt\,\left(\frac{1}{2}\, \dot q^2+V(q)\right)\ ,
\end{equation}
and its naive discretization equals
\begin{equation}
S_N=\sum_{n=0}^{N-1}\left(\frac{\delta^2_n}{2\epsilon_N}+\epsilon_NV(\bar q_n)\right)\ ,
\end{equation}
where $\delta_n=q_{n+1}-q_n$, and $\bar q_n=\frac{1}{2}(q_{n+1}+q_n)$. Note that we are using units in which the particle mass and $\hbar$ have been set to unity and that we are evaluating path integrals in the so-called mid-point ordering prescription.

\begin{figure}[!ht]
\centering
\includegraphics[width=13cm]{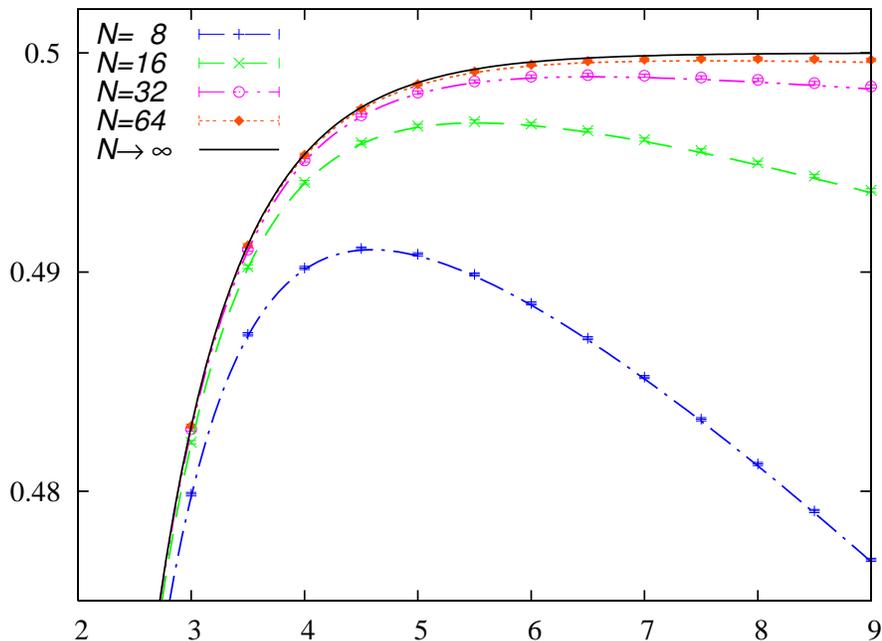}
\caption{The curves depict the exact solution of the discretized free energies $F_N(\beta)$ for the harmonic oscillator in the left ordering prescription given in Eq.~(\ref{harmonic}) for various values of $N$. The data points give the results and error bars of the corresponding numerical calculations, used to verify the code. Parameters are $\omega=1$ and $N_{MC}=10^7$.}\label{exact}
\end{figure}

From the above we have obtained a path integral representation for the partition function that is directly amenable to numerical evaluation. On the other hand, by evaluating the trace in Eq.~(\ref{partition}) in the energy basis we find
\begin{equation}
Z(\beta)\equiv e^{-\beta F(\beta)}=\sum_{n=0}^\infty e^{-\beta E_n}\ .
\end{equation}
As we can see, the partition function, or equivalently the free energy $F(\beta)$, completely determines the energy spectrum and vice-versa. For example, if we define a series of auxiliary functions as
\begin{equation}
\label{auxiliary}
F_n(\beta)=-\frac{1}{\beta}\ln\left( e^{-\beta F}-\sum_{i=0}^{n-1}e^{-\beta E_i}\right)\ ,
\end{equation}
then it immediately follows that $F_n(\beta)\to E_n$ for large $\beta$. It would be ideal, therefore, if we could calculate the free energy (and the other auxiliary functions) for arbitrarily large values of $\beta$. This is not possible in numerical calculations. First of all the calculations become much more demanding with growth of ``time of propagation" $\beta$ (just as the physics becomes more interesting). More importantly, when doing numerical calculations we evaluate discretized quantities such as $F_N$, and the $N\to\infty$ and $\beta\to\infty$ limits that one would need to perform do not commute. The best way to see this is to look at the free energy of an exactly solvable model -- the harmonic oscillator. In this case the $N$-fold discretized free energy (in the left ordering prescription) equals \cite{kleinert}
\begin{equation}
\label{harmonic}
F_N(\beta)=\frac{1}{\beta}\ln\left(2\sinh\left(\tilde{\omega}\beta\right)\right),
\end{equation}
where $\tilde{\omega}=(2/\epsilon_N)\,\mathrm{arc sinh}(\omega\epsilon_N/2)$. This solution is illustrated in Fig.~\ref{exact}. It follows that, unlike its continuum limit $F(\beta)$, the discretized free energy $F_N(\beta)$ does not tend to a constant value for large $\beta$. Said another way, the discretized energy levels themselves depend on $\epsilon_N$ and thus on $\beta$. For example, for the harmonic oscillator we have $E_{N,n}(\epsilon_N)=\tilde{\omega}(n+1/2)$.

In the case of a general theory the free energy is related to its discretized value as $F(\beta)-F_N(\beta)=\mathrm{O}(\epsilon_N)$. We see that $F_N(\beta)$ slowly converges to its continuum limit, i.e. that we need a large number of discretization points $N$ to approach that value. In addition, the larger the value of $\beta$ we want, the larger $N$ must be in order to achieve a given accuracy. The price we pay is in the computer time which grows linearly with $N$.

\begin{figure}[!ht]
\centering
\includegraphics[width=13cm]{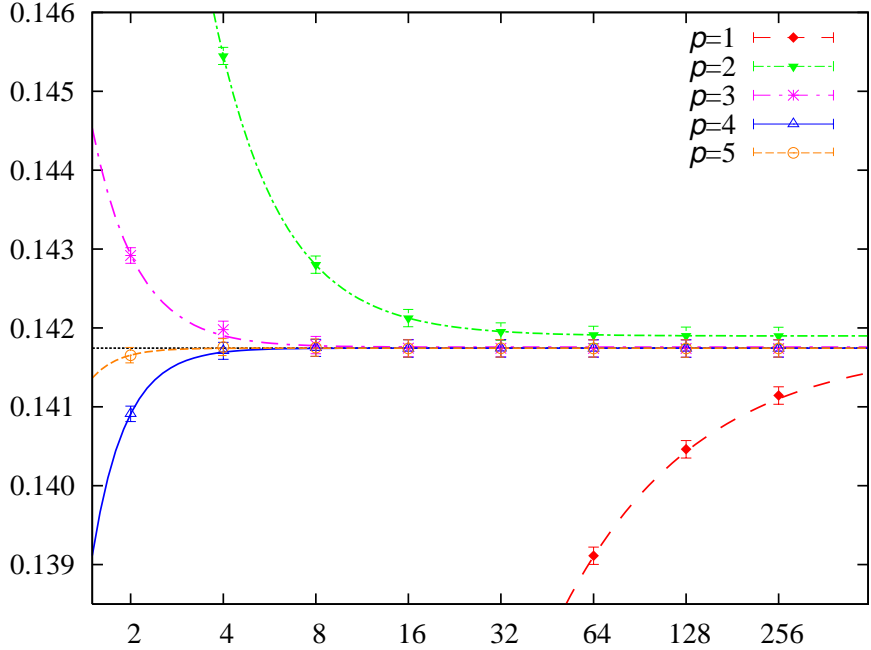}
\caption{The dependance of $F_N^{(p)}(\beta)$ on $N$ for different levels $p$. The plot is for the anharmonic oscillator with quartic coupling $g=1$, inverse temperature $\beta=1$ and $N_{MC}=10^7$ Monte Carlo samples. The same kind of behavior is seen for other parameters as well as for other potentials.
\label{FpN}}
\end{figure}

A recent series of papers \cite{prl,prb,euler} analytically studied the relation between path integral discretizations of different coarseness for the case of a general theory. This work resulted in a systematic construction of a hierarchy of $N$-fold discretized effective actions $S^{(p)}_N$ labeled by a whole number $p$ and built up from the naively discretized action in the mid-point prescription (corresponding to $p=1$).
The level $p$ effective action leads to discretized transition amplitudes and expectation values differing from the continuum limit by a term of order $1/N^p$. Thus, moving up the hierarchy we are guaranteed to get expressions which converge ever faster to the continuum limit. The direct application of these results to the free energy gives
\begin{equation}
F(\beta)-F_N^{(p)}(\beta)=\mathrm{O}(\epsilon_N^p)\ .
\label{converge}
\end{equation}
For a given inverse temperature $\beta$, and for $\epsilon_N\lesssim 1$ the discretized free energy $F^{(p)}_N(\beta)$ converges faster to the continuum as we increase the hierarchy level $p$. This is illustrated in Fig.~\ref{FpN}.

When using the path integral Monte Carlo method to calculate the free energy $F(\beta)$ there are two sources of errors. The first comes from the limited number of Monte Carlo samples $N_{MC}$ and is proportional to $N_{MC}^{-1/2}$. The second type of error comes from discretization -- in our case from approximating the free energy with $F_N^{(p)}(\beta)$ for some $N$ and $p$. As we have seen, for a given $\beta$ this discretization error is proportional to $N^{-p}$. These two types of errors should optimally be of the same order, e.g. there is no point in decreasing the discretization error bellow the Monte Carlo error as this would not decrease the overall error. In practice we fix the precision we want by choosing the number of Monte Carlo samples and then decrease the discretization error to match this either by increasing $N$ or the hierarch level $p$. The second choice is far better; however, since computation times grow linearly with $N$, but are almost independent of $p$ (at least for $p\le 9$, the hierarchy levels studied in \cite{prl,prb}). As a consequence of this, the speedup coming from using higher values of $p$ at fixed precision $\delta$ is proportional to $\delta^{-1+1/p}$. Therefore, by using $p=9$ we are in fact quite near to the point of optimal benefit for which the speedup of the new method is inversely proportional to the precision. As an illustration, for two decimal precision the new method gives a hundred fold speedup over the defining algorithm, for four decimal precision the speedup is ten thousand fold, etc. It is important to note that the greatest utility of the new evaluation scheme is, therefore, when calculating quantities with high precision. We stress that all of this holds for $\epsilon_N\lesssim 1$, i.e. as long as $N\gtrsim\beta$ is satisfied.

\section{Numerical Results}

As we have seen in the previous section, $F(\beta)$ can be evaluated with arbitrary precision on any interval of inverse temperatures $[0,\beta_{max}]$ for any given potential by appropriately increasing and adjusting $N$, $p$, and $N_{MC}$. Let us now numerically compare the quality of different discretizations of the free energy $F_N^{(p)}$ with $F^*$, the most accurate one that may be calculated on a given set $\{\beta_i\}$. To do this we use the standard $\chi^2$ function,
\begin{equation}
\chi^2\left(N,p\right)=\frac{1}{M}\sum_{\{\beta_i\}} \frac{\left( F_N^{(p)}(\beta_i)-F^*(\beta_i)\right)^2}
{\left(\Delta F_N^{(p)}(\beta_i)\right)^2+\Big(\Delta F^*(\beta_i)\Big)^2}\ ,
\end{equation}
where $M$ is the number of points in the set $\{\beta_i\}$, and $\Delta F$ is the Monte Carlo error. By including the Monte Carlo error of $F^*$ into the $\chi^2$ weights we took into account the fact that it is also calculated numerically. $\chi^2$ should be around one for well optimized $N$ and $p$. Note that $\chi^2\gg 1$ if the exact value of $F^*$ is not within the error bars of $F_N^{(p)}$, while $\chi^2\ll 1$ if the Monte Carlo error is too large.

\begin{figure}[!ht]
\centering
\includegraphics[width=13cm]{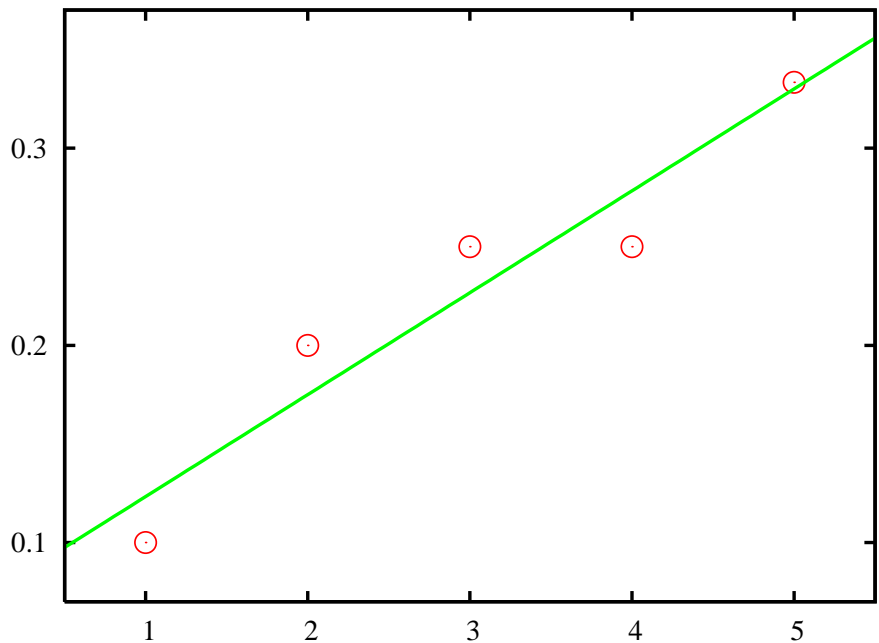}
\caption{Pairs of N and p which give similar values od $\chi^2$. The plot gives $1/\log_2 N$ on $y$ axis as a function of $p$. The general behavior is illustrated on the case of the anharmonic oscillator with quartic coupling $g=1$, $\beta_{max}=8$, $N_{MC}=10^6$, $\chi^2\approx 2 - 4$.\label{lnNvsp}}
\end{figure}

We conducted this test on the anharmonic oscillator with quartic coupling $V(q)=\frac{1}{2}q^2+\frac{g}{4!}q^4$. The discretized free energies were calculated for $\beta\in [0.5,8]$ with step $0.5$, $N\le 1024$ and $p=1,2,\ldots,9$. The number of Monte Carlo samples used was $10^6$. The comparisons were done for a range of coupling constants $g\in\{0,0.1,1,10,100,1000\}$. Taking $F_{1024}^{(9)}$ as the exact result, we calculated $\chi^2$ for each pair of parameters $(N,p)$ and coupling $g$, and looked for $(N,p)$ pairs with approximately the same values of $\chi^2$. These pairs are given in Fig.~\ref{lnNvsp}. As we can see, the relation $1/\log_2 N\propto p$ that is implicit in Eq.~(\ref{converge}) actually holds, i.e. the error indeed scales as $N^{-p}$.

\begin{figure}[!b]
\centering
\includegraphics[width=13cm]{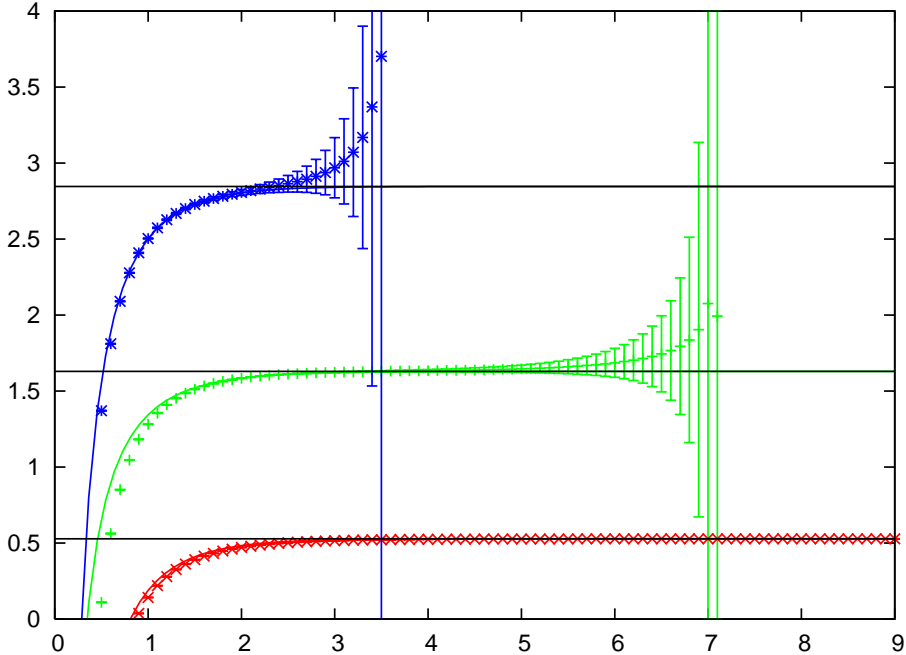}
\caption{Dependance of the free energy $F$ and the associated auxiliary functions $F_1$ and $F_2$ on $\beta$ for the anharmonic oscillator with quartic coupling $g=1$. The solid lines are the fits to curves of the form given in Eq.~(\ref{fit}). The horizontal lines in black correspond to the energy levels $E_n$ determined from these fits (see Table~\ref{tableaho}). Numerical simulations were performed with $p=9$ level improved actions, $N=256$, and $N_{MC}=10^7$.\label{Fbeta}}
\end{figure}

We now turn to calculating the energy spectrum using the outlined efficient procedure for evaluating the free energy of a general theory. For the range of inverse temperatures $\beta$ that will be used for numerical calculations of the energies we choose $\beta_{max}$ so that $F_{N}(\beta)=F(\beta)$ within the error bars on the whole $[0,\beta_\mathrm{max}]$ interval. We also need to ensure that all the assumptions mentioned above hold ($\epsilon_N\lesssim 1$, $\beta_{max}$ fixed). The free energy $F(\beta)$ and all its auxiliary functions can be written as
\begin{equation}
F_n(\beta)=E_n-\frac{1}{\beta}\ln\left(1+\sum_{i=n+1}^{\infty}e^{-\beta\left(E_i-E_n\right)}\right)\ .
\end{equation}
As a result, we have fit the numerical data to functions of the form
\begin{equation}
\label{fit}
F_n(\beta)=E_n-\frac{1}{\beta}\ln\left(1+A\,e^{-B\beta }\right)\ ,
\end{equation}
where $E_n$, $A$ and $B$ are the parameters of the fit. Fig.~\ref{Fbeta} shows the free energy $F(\beta)$ (approximated by its discretization for $N=256$ and $p=9$) along with the associated auxiliary functions $F_1(\beta)$, and $F_2(\beta)$ for the anharmonic oscillator with quartic coupling $g=1$. Note that the class of functions given in Eq.~(\ref{fit}) gives a better fit for larger values of $\beta$. This can indeed be explicitly seen from Fig.~\ref{Fbeta}.
The data points for the free energy $F(\beta)$ were obtained directly from our Monte Carlo simulations and were used to determine the ground state energy $E_0$. The auxiliary functions $F_n(\beta)$ were obtained recursively using Eq.~(\ref{auxiliary}) and the already determined energy levels. The error bars presented in the figure also follow directly from Eq.~(\ref{auxiliary}) and are given by
\begin{equation}
\label{errors}
\Delta F_n=\frac{\Delta F e^{-\beta F}+\sum_{i=0}^{n-1}\Delta E_ie^{-\beta E_i}}{e^{-\beta F} -\sum_{i=0}^{n-1}e^{-\beta E_i}}\ .
\end{equation}
For large inverse temperatures $\beta$ the above denominator becomes exponentially small, and so the error bars become very large. Such points soon cease to give relevant contributions to the calculations of the corresponding energy level owing to the fact that we use a weighted fit. Note that, in fact, the lack of exponential growth of error bars with $\beta$ is an indication of bad data points!

This effect of growing error bars becomes more pronounced for higher energy levels. In addition, from Eq.~(\ref{errors}) we see that there is an accumulation of errors associated with all the lower energy levels. Both of these effects taken together give practical limits to the number of energy levels we can calculate. The precise depth to which we can probe the energy spectrum depends on the number of Monte Carlo samples used as well as the number of points $\beta_i$ selected within the range of inverse temperatures available to us. As an illustration, Table~\ref{tableaho} gives the low lying energy levels of the anharmonic oscillator for several values of coupling $g$. For all of these calculations we use the same range of $\beta$. The ground state energy level was calculated to five significant digits for all values of $g$. As we have already noted the errors increase as we go to higher energy levels. In fact, this increase is faster for larger couplings since then the energies themselves become higher and so the $e^{-\beta E_n}$ terms become much smaller.
\begin{table}[!ht]
\begin{center}
\centering
\begin{tabular}{|r|c|c|c|c|}\hline
$g$ & $E_0$ & $E_1$ & $E_2$ & $E_3$ \\\hline\hline
0&0.49993(2)& 1.502(2)&2.48(6)&3.6(5)\\\hline
0.1& 0.50301(2)&1.516(1)&2.54(5)&3.5(2)\\\hline
1& 0.52765(2)&1.6295(8)&2.85(2)&3.98(7)\\\hline
10& 0.67335(2)&2.230(1)&4.12(2)&\\\hline
100& 1.16247(4)&4.058(6)&&\\\hline
1000& 2.3578(2)&&&\\\hline
\end{tabular}
\end{center}
\vspace{0.2cm}
\caption{Low lying energy levels of the anharmonic oscillator with quartic coupling $g$, calculated using $N=256$, $p=9$, and $N_{MC}=10^7$.\label{tableaho}}
\end{table}

\begin{table}[!ht]
\begin{center}
\centering
\begin{tabular}{|c|c|c|c|c|c|}\hline
$\alpha$ & $\lambda$ & $E_0$ & $E_0^{exact}$ & $E_1$ & $E_1^{exact}$ \\\hline\hline
0.25&5.5& -0.6329(2)&-0.63281& -0.3819(7)&-0.38281\\\hline
0.25&15.5& -6.5704(6)&-6.57031&-5.694(9)&-5.69531\\\hline
0.5&5.5& -2.5313(3)&-2.53125&-1.530(3)&-1.53125\\\hline
0.5&15.5& -26.281(1)&-26.2813&-22.80(3)&-22.7813\\\hline\hline\hline
$\alpha$ & $\lambda$ & $E_2$ & $E_2^{exact}$ &$E_3$ & $E_3^{exact}$ \\\hline\hline
0.25&5.5&-0.18(2)&-0.19531&-0.09(3)&-0.07031\\\hline
0.25&15.5&-4.92(2)&-4.88281&-3.8(4)&-4.13281\\\hline
0.5&5.5&-0.80(2)&-0.78125&-0.31(6)&-0.28125\\\hline
0.5&15.5&-19.6(5)&-19.5313&-16.9(9)&-16.5313\\\hline
\end{tabular}
\end{center}
\vspace{0.2cm}
\caption{Low lying energy levels of the modified P\"oschl-Teller potential, calculated using $N=256$, $p=9$, and $N_{MC}=10^7$.\label{tablempt}}
\end{table}

We have conducted explicit Monte Carlo calculations of the spectra of the P\"oschl-Teller and Morse potentials and have obtained the same qualitative behavior. In particular, we have explicitly determined that the expected speedup in convergence, coming from using the $p$-level hierarchy of effective actions, holds for all of these potentials.

Obtained low lying energy levels for several values of the parameters of the modified P\"oschl-Teller potential,

\begin{equation}
V(q)=-\frac{\alpha^2}{2}\frac{\lambda(\lambda-1)}{\cosh^2 \alpha x}\ ,
\end{equation}

are given in Table~\ref{tablempt}. We considered this exactly solvable potential since it allows comparison of numerically calculated energy levels and the exact ones, given by
$$E_n^{exact}=-\frac{\alpha^2}{2}(\lambda - 1 - n)^2\ ,\, 0\leq n\leq \lambda - 1\ ,\, n\in\mathbb{N}\ .$$
As can be seen from Table~\ref{tablempt}, numerical results are in excellent agreement with the exact energy levels even for a small value of discretization coarseness $N$.

As a conclusion, we have investigated a newly developed method for increasing the convergence of path integrals to the continuum limit. The method has previously been shown to lead to a many order of magnitude speedup in the numerical evaluation of path integrals for transition amplitudes \cite{prl,prb,euler} and expectation values \cite{estimators}. In this paper we have applied that method to the evaluation of energy spectra. We have shown that the above stated increase in convergence leads to a significant increase of the efficiency of path integral Monte Carlo calculations of low lying energy levels of a generic theory. The analytical results were checked explicitly in a series of Monte Carlo simulations of several distinct models over a wide range of parameters.


\begin{thebibliography}{00}

\bibitem{prl}
A. Bogojevi\'c, A. Bala\v z, and A. Beli\'c, Phys. Rev. Lett. {\bf 94}, 180403 (2005).
\bibitem{prb}
A. Bogojevi\'c, A. Bala\v z, and A. Beli\'c, Phys. Rev. B {\bf 72}, 064302 (2005).
\bibitem{euler}
A. Bogojevi\'c, A. Bala\v z, and A. Beli\'c, Phys. Lett. A {\bf 344}, 84 (2005).
\bibitem{estimators}
J. Gruji\'c, A. Bogojevi\'c, and A. Bala\v z, submitted to Phys. Lett. A
\bibitem{feynmanhibbs}
R. P. Feynman and A. R. Hibbs,
\emph{Quantum Mechanics and Path Integrals}
(McGraw-Hill, New York, 1965).
\bibitem{feynman}
R. P. Feynman,
\emph{Statistical Mechanics}
(W. A. Benjamin, New York, 1972).
\bibitem{itzyksonzuber}
C. Itzykson and J.-B. Zuber,
\emph{Quantum Field Theory}
McGraw-Hill, New York, 1980.
\bibitem{coleman}
S. Coleman,
\emph{Aspects of Symmetry}
Cambridge University Press, 1985.
\bibitem{itzyksondrouffe}
C. Itzykson and J.-M. Drouffe,
\emph{Statistical Field Theory}
(Cambridge University Press, 1991).
\bibitem{parisi}
G. Parisi,
\emph{Statistical Field Theory}
(Addison Wesley, New York, 1988).
\bibitem{barkerhenderson}
J. A. Barker and D. Henderson,
Rev. Mod. Phys. {\bf 48}, 587 (1976).
\bibitem{pollockceperley}
E. L. Pollock and D. M. Ceperley,
Phys. Rev. B {\bf 30}, 2555 (1984).
\bibitem{ceperley}
D. M. Ceperley,
Rev. Mod. Phys. {\bf 67}, 279 (1995).
\bibitem{kleinert}
H. Kleinert,
\emph{Path Integrals in Quantum Mechanics, Sta\-tistics, Polymer Physics, and Financial Markets}
(World Scientific, 2004).
\bibitem{jaggedness}
A. Bogojevic, A. Balaz, and A. Belic, Phys. Lett. A {\bf 345}, 258 (2005).
\bibitem{egee}
http://www.eu-egee.org/
\bibitem{seegrid}
http://www.see-grid.eu/
\bibitem{scl}
http://scl.phy.bg.ac.yu/speedup/


\end{thebibliography}
\end{document}